\documentclass[final, conference, twocolumn]{IEEEtran}
\IEEEoverridecommandlockouts

\usepackage[pdftex]{graphicx}
\usepackage{bm}
\usepackage{cite}
\usepackage{amsmath,amssymb,amsfonts}
\usepackage{mathtools}
\usepackage{cleveref}
\usepackage{autonum}
\usepackage[subrefformat=parens]{subcaption}

\crefname{figure}{Fig.}{Figs.}
\crefname{table}{Table}{Tables}
\crefname{algorithm}{Algorithm}{Algorithms}
\crefname{section}{section}{sections}
\crefname{subsection}{section}{sections}
\crefname{subsubsection}{section}{sections}
\crefname{equation}{}{}

\newcommand{\brak}[1]{\displaystyle \left( {#1} \right)}
\newcommand{\floor}[1]{\displaystyle \left \lfloor {#1} \right \rfloor}
\newcommand{\round}[1]{\displaystyle \left[ {#1} \right]}

\newcommand{\bi}{\bm{i}}
\newcommand{\bj}{\bm{j}}
\newcommand{\bfo}{f}

\newcommand{\bfbf}{f_{BF}}
\newcommand{\fgf}{f_{GF}}
\newcommand{\fti}{f_{TI}}
\newcommand{\bfbg}{f_{BG}}
\newcommand{\sigr}{\sigma_r}
\newcommand{\sigs}{\sigma_s}

\newcommand{\sigg}{\sigma_g}

\newcommand{\bfv}{\bm{fv}}
\newcommand{\bv}{\bm{v}}
\newcommand{\bw}{\bm{w}}
\newcommand{\rr}{r \sigr / \sigs}
\newcommand{\gridds}{grid^{2D}}
\newcommand{\gf}{grid_f}
\newcommand{\gfds}{\gf^{2D}}

\newcommand{\eround}[1]{\round{\begin{array}{@{}c@{}} {#1} \end{array}}}
\newcommand{\am}{\mspace{-3mu} - \mspace{-3mu}}
\newcommand{\ap}{\mspace{-3mu} + \mspace{-3mu}}
\newcommand{\aeq}{\mspace{-3mu} = \mspace{-3mu}}
\newcommand{\at}{\mspace{-3mu} \times \mspace{-3mu}}
\newcommand{\ad}{\mspace{-3mu} \textrm{/} \mspace{-3mu}}
\newcommand{\ape}{\mspace{-3mu} \mathrel{{+}{=}} \mspace{-3mu}}
\newcommand{\asum}[1]{\mspace{-3mu} \sum_{#1} \mspace{-5mu}}
\newcommand{\ii}{\textrm{II} \aeq 1}

\let\%\relax
\DeclareFontFamily{OT1}{pxr}{}
\DeclareFontShape{OT1}{pxr}{m}{n}{<->pxr}{}
\DeclareSymbolFont{letA}{OT1}{pxr}{m}{n}
\DeclareMathSymbol{\%}{0}{letA}{`\%}

\usepackage[noend]{algpseudocode}
\usepackage{algorithm}

\def\BibTeX{{\rm B\kern-.05em{\sc i\kern-.025em b}\kern-.08em
    T\kern-.1667em\lower.7ex\hbox{E}\kern-.125emX}}
\begin{document}

\title{An FPGA-Based Fully Pipelined Bilateral Grid \\ for Real-Time Image Denoising}

\author{
\IEEEauthorblockN{Nobuho Hashimoto\IEEEauthorrefmark{1}, Shinya Takamaeda-Yamazaki\IEEEauthorrefmark{2}}
\IEEEauthorblockA{
\textit{The University of Tokyo}\\
\IEEEauthorrefmark{1}hashimoto-nobuho949@g.ecc.u-tokyo.ac.jp, \IEEEauthorrefmark{2}shinya@is.s.u-tokyo.ac.jp}
}


\maketitle

\begin{abstract}
The bilateral filter (BF) is widely used in image processing because it can perform denoising while preserving edges. It has disadvantages in that it is nonlinear, and its computational complexity and hardware resources are directly proportional to its window size. Thus far, several approximation methods and hardware implementations have been proposed to solve these problems. However, processing large-scale and high-resolution images in real time under severe hardware resource constraints remains a challenge.

This paper proposes a real-time image denoising system that uses an FPGA based on the bilateral grid (BG). In the BG, a 2D image consisting of x- and y-axes is projected onto a 3D space called a ``grid,'' which consists of axes that correlate to the x-component, y-component, and intensity value of the input image. This grid is then blurred using the Gaussian filter, and the output image is generated by interpolating the grid. Although it is possible to change the window size in the BF, it is impossible to change it on the input image in the BG. This makes it difficult to associate the BG with the BF and to obtain the property of suppressing the increase in hardware resources when the window radius is enlarged.

This study demonstrates that a BG with a variable-sized window can be realized by introducing the window radius parameter wherein the window radius on the grid is always 1. We then implement this BG on an FPGA in a fully pipelined manner. Further, we verify that our design suppresses the increase in hardware resources even when the window size is enlarged and outperforms the existing designs in terms of computation speed and hardware resources.
\end{abstract}


\section{Introduction}
The bilateral filter (BF) is popularly used as an edge-preserving smoother in many image processing applications such as tone mapping \cite{tone_mapping}, stylization \cite{stylization}, upsampling \cite{upsampling}, and optical-flow estimation \cite{optical_flow}. One of the most important applications is medical image denoising \cite{medical_denoising}. \cite{denoising_quality} demonstrated that the quality of medical images such as X-Ray and CT significantly improved after they were processed using the BF. Although this procedure demands real-time responses for interactive operations, real-time processing is still difficult under the severe constraints of hardware resources for large-scale and high-resolution images.

Considering a two-dimensional 8-bit grayscale image $\bfo: \Omega \to \mathbb{I}$ where $\Omega \subset \mathbb{N}^2$ is the domain of the image and $\mathbb{I} \aeq \{i \mid i \aeq 0, 1, \ldots, 255 \}$ is the intensity range, the BF output $\bfbf: \Omega \to \mathbb{I}$ is given by
\begin{align}
    \includegraphics[width=0.91\linewidth]{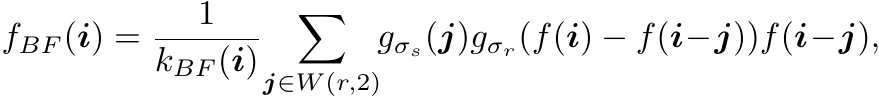}
    \label{bf}
\end{align}
where $k_{BF}: \Omega \to \mathbb{R}$ represents the normalization term
\begin{align}
    k_{BF}(\bi) = \asum{\bj \in W(r, 2)} g_{\sigs}(\bj) g_{\sigr}(\bfo(\bi) - \bfo(\bi \am \bj)),
\end{align}
$W(r, d) \aeq [-r, r]^d$ represents a hypercube around the pixel of interest; $r \in \mathbb{N}$ and $d \in \mathbb{N}$ denote the window radius and dimension, respectively; and $g_{\sigs}: \mathbb{I}^2 \to \mathbb{R}$ and $g_{\sigr}: \mathbb{I} \to \mathbb{R}$ represent the Gaussian spatial and range kernels, respectively. Hereafter, $g_{\sigma}$ denotes the Gaussian function, where $\sigma$ is the standard deviation.

The range kernel provides edge-preserving characteristics; however, it makes the BF nonlinear, which causes difficulty in acceleration. If the BF is linear, we can easily apply the methods presented in \cite{rec_gaus1,rec_gaus2}. Therefore, some approaches have attempted to remove nonlinearity by quantizing the range space \cite{real_time_O1,constant_time_median,clustering}, and by approximating the range kernel using trigonometric functions \cite{shiftability1,shiftability2}, the Taylor polynomials \cite{constant_time_O1,fast_and_provably}, and the DFT (Discrete Fourier Transform) \cite{dft}. Other approaches to acceleration also exist, wherein input images are projected onto the other smaller space such as the histogram \cite{histogram_Olog}, bilateral grid (BG) \cite{bilateral_grid}, and adaptive manifolds \cite{adaptive_manifolds}. These approaches are often implemented on a GPU, but they are not as small-scale and energy-efficient as FPGA implementations \cite{energy_consumption}. Therefore, brute force FPGA implementations are proposed \cite{brute_force_mapping,brute_force_polynomial,fully_synchronized}. The aforementioned approaches \cite{shiftability1,fast_and_provably,histogram_Olog} have also been implemented on an FPGA \cite{co-design,reconfigurable,histogram_O1}. However, these implementations still have at least one of the following unsolved problems: low throughput, large memory footprint, and an increase in hardware resources depending on the window radius.

To solve these problems, this paper proposes a novel method for the BG \cite{bilateral_grid} and its fast and small FPGA implementation for the BF on grayscale images. Here, the BG can suppress the increase in hardware resources, and an FPGA can achieve fast and small-scale implementation. Therefore, we realized a high throughput, low latency accelerator with a low memory footprint using noniterative and sequential processing. The major contributions of this study are summarized as follows.
\begin{enumerate}
    \item The BG is enhanced so that the window size of input images can be varied.
    \item The fully pipelined FPGA implementation is proposed for the proposed BG so that it can suppress the increase in the hardware resources.
    \item The proposed design is implemented on an actual FPGA board, and it outperforms the other existing designs in terms of computation speed and hardware resources.
\end{enumerate}


\section{FPGA-based Bilateral Grid \\with a Variable-Sized Window} \label{sec:proposed_method}
\subsection{Bilateral Grid with a Variable-Sized Window}
First, we attempt to change the BG window radius on an input image. The existing BG \cite{bilateral_grid} does not consider the window radius on the input image, but on the grid. This makes it difficult to associate the BG with the BF, and to suppress the increase in hardware resources when the window radius is enlarged, because if the radius on the grid increases, the resources will increase in the same way as the original BF. By following the derivation of the existing BG, the output of the BG with a variable-sized window $\bfbg: \Omega \to \mathbb{I}$ is obtained by rewriting \cref{bf} as
\begin{align}
    \includegraphics[width=0.91\linewidth]{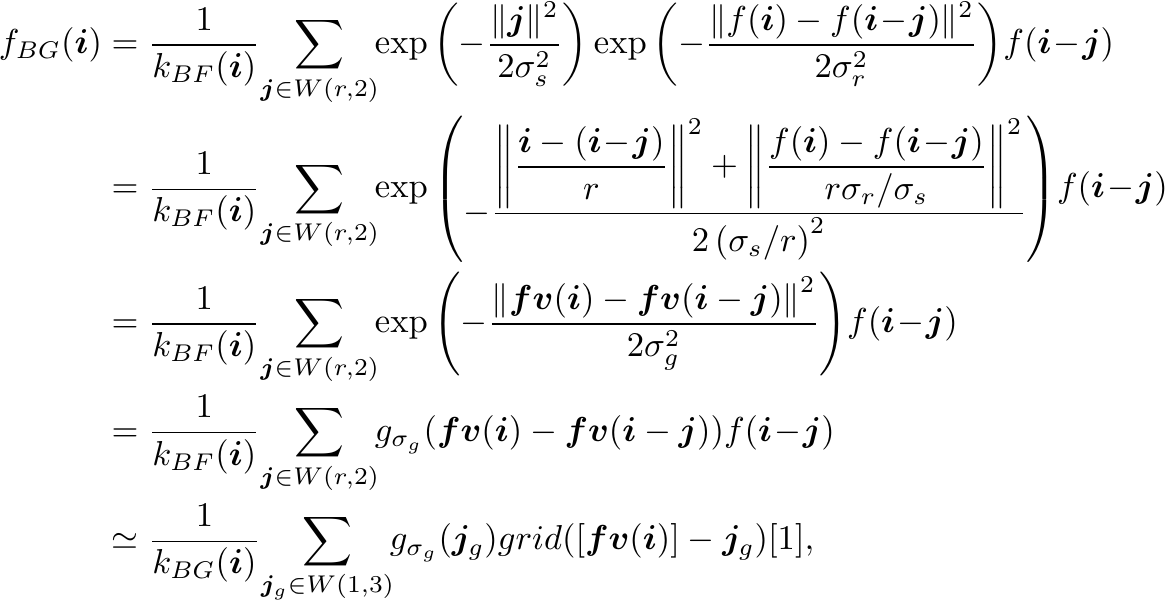}
    \label{bgwvsw}
\end{align}
where $\sigg \aeq \sigs / r$ denotes the standard deviation, $k_{BG}: \Omega \to \mathbb{R}$ denotes the normalization term
\begin{align}
    k_{BG}(\bi) = \asum{\bj_{g} \in W(1, 3)} g_{\sigg}(\bj_{g}) grid(\round{\bfv(\bi)} - \bj_{g})[0],
\end{align}
$\bfv: \Omega \to \Delta (\subset \mathbb{R}^3)$ represents the feature vector
\begin{align}
    \bfv(\bi) = \brak{\dfrac{\bi}{r}, \dfrac{f(\bi)}{\rr}},
    \label{feature_vector}
\end{align}
and $grid: \Gamma \to \mathbb{N}^2$ expresses the grid space; $\Gamma \subset \mathbb{N}^3$ represents the set of lattice points in the grid, the second term of $grid$ is the sum of the intensity values in the elements, and the first term is the number of pixels present. Here, the window radius in the last line in \cref{bgwvsw} is fixed at 1 to make the computation of the filtering easier and faster. To achieve this, while the output is maintained to be similar to the BF, all pixels in the BF window need to be included in the window of the proposed BG. Thus, we introduce $r$, as shown in the second line in \cref{bgwvsw}.

Then, the algorithm of the proposed BG (the last line in \cref{bgwvsw}) is described as
\begin{align}
    \quad \includegraphics[width=\linewidth-10pt]{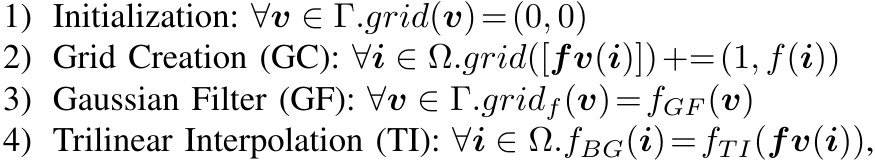}
\end{align}
where $\fgf: \Gamma \to \mathbb{R}$ represents the GF around the element of interest in the grid, $\gf: \Gamma \to \mathbb{R}$ denotes the grid after the GF, and $\fti: \Delta \to \mathbb{I}$ denotes the TI for the elements on $\gf$. Here, the function $\fgf: \Gamma \to \mathbb{R}$ is given by
\begin{align}
    \includegraphics[width=0.91\linewidth]{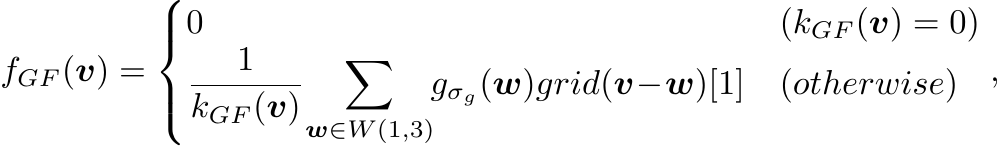}
    \label{gf3D}
\end{align}
where $k_{GF}: \Gamma \to \mathbb{R}$ denotes the normalization term
\begin{align}
    k_{GF}(\bv) = \asum{\bw \in W(1, 3)} g_{\sigg} (\bw) grid(\bv \am \bw)[0].
    \label{gf3D_norm}
\end{align}
Then, the function $\fti: \Delta \to \mathbb{I}$ is given by
\begin{align}
    \fti(\bm{p}) = \eround{\displaystyle \sum_{i, j, k \in \{0,1\}} \mspace{-5mu} x_i y_j z_k \cdot \gf(\lfloor \bm{p} \rfloor +(i, j, k))},
    \label{interpolate}
\end{align}
where $x_i$, $y_j$, $z_k$ denote the coefficients
\begin{align}
    (x_i, y_j, z_k) = |\bm{p} - \floor{\bm{p}} - (i, j, k)|.
\end{align}

\subsection{Overall Accelerator Architecture}
\begin{figure}[tbp]
    \centering
    \includegraphics[width=0.9\linewidth,pagebox=mediabox]{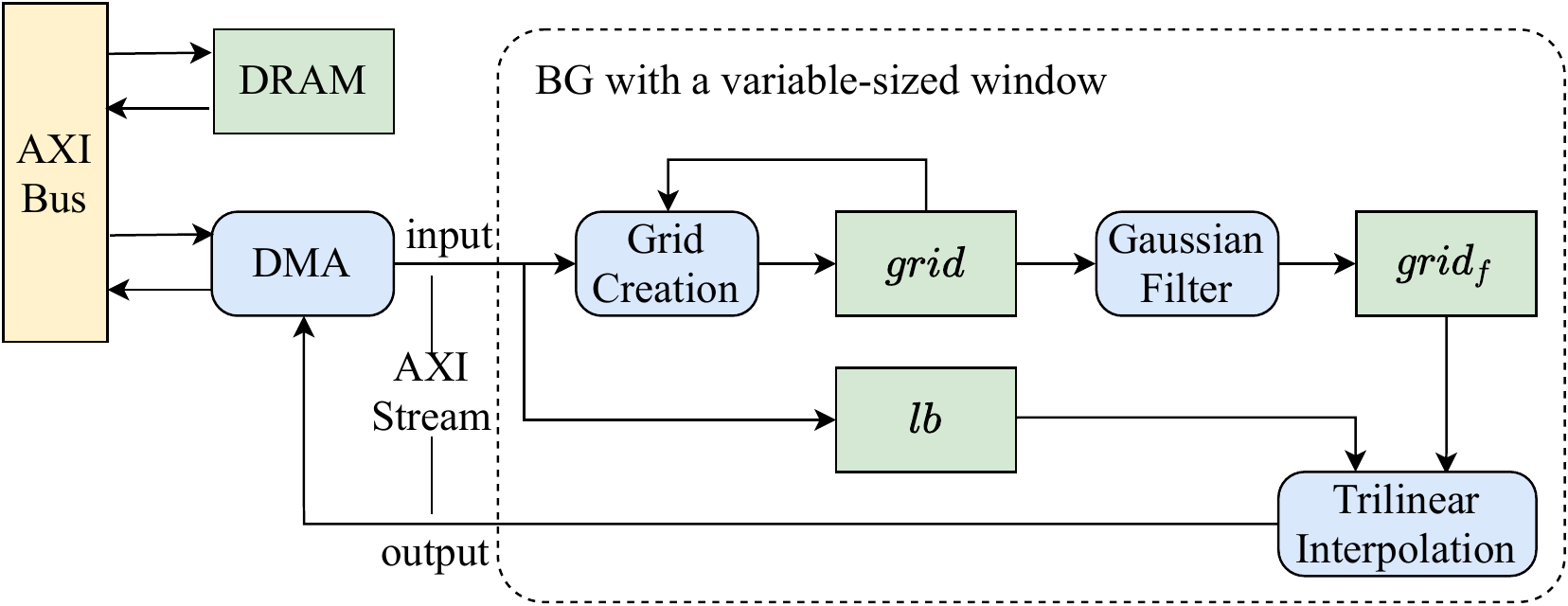}
    \caption{Overall accelerator architecture and data flow of the FPGA-based BG with a variable-sized window.}
    \label{fig:overall}
\end{figure}
The overall architecture of the FPGA-based BG with a variable-sized window is shown in \cref{fig:overall}. First, the input image pixels $(x, y)$ are read one by one from the DRAM using the AXI bus and DMA. The GC then converts the input image into the grid when each input pixel is read; therefore, the grid elements $(x, y, z)$ are filled in line by line. The output values of this operation are stored in BRAMs (Block RAMs) $grid$. Here, the values in $grid$ must be read for the GC because the operation $grid([\bfv(\bi)]) \ape (1, f(\bi))$ is performed. The GF then blurs the grid after a bare minimum of elements ($3 \at 3 \at 3$ cube around the element of interest) are prepared. Thus, it blurs one plane in the $r$ lines of the input image. The output values of this operation are stored in BRAMs $\gf$. Because the input of the TI is a feature vector \cref{feature_vector} of the pixel of interest, the input image must be stored in BRAMs $lb$ (line buffer). Finally, the TI is executed after a bare minimum of elements (eight nearest elements around the point of interest) are blurred by the GF. Therefore, $r$ lines of the output image are obtained per $r$ lines of the input image.

\subsection{Hardware Optimization}
Here, we focus on a $w \at h$ 8-bit grayscale image. For the sake of simplicity, this paper does not explain corner cases in detail, such as the leftmost and rightmost lines; however, in essence, these can be implemented similarly to the other cases.

As derived from \cref{feature_vector,interpolate}, the domain of the grid is defined as
\begin{align}
    \includegraphics[width=\linewidth]{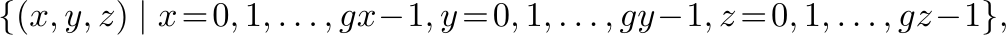}
\end{align}
where the constants $gx$, $gy$, and $gz$ denote
\begin{align}
    (gx, gy, gz) = \brak{\floor{\dfrac{h}{r}} + 2, \floor{\dfrac{w}{r}} + 2, \floor{\dfrac{255}{\rr}} + 2}.
    \label{grid_size}
\end{align}

To separate the input image, we define $gg$ and $gi$ as
\begin{align}
    gg(x, y) &= \left\{ (ix, iy) \mspace{1mu} \left| \mspace{1mu} \brak{ \round{\dfrac{ix}{r}}, \round{\dfrac{iy}{r}} } = (x, y) \right. \right\} \\
    gi(x, y) &= \left\{ (ix, iy) \mspace{1mu} \left| \mspace{1mu} \brak{ \floor{\dfrac{ix}{r}}, \floor{\dfrac{iy}{r}} } = (x, y) \right. \right\},
\end{align}
so that all pixels in $gg(x, y)$ are projected onto the $grid$ elements with the same x- and y-components $grid(x, y, \ast)$ and all pixels in $gi(x, y)$ require the $\gf$ elements with the same x- and y-components $\gf(x, y, \ast)$ for the TI. Hereafter, the notation asterisk $\ast$ is used as a wildcard; for example, $grid(x, y, \ast)$ denotes $\{ grid(x, y, z) \mid z \aeq 0, 1, \ldots, gz \am 1 \}$.

The overall pseudo code is shown in \cref{alg:bg}, where $L_1$, $L_2$, and $L_3$ are LUTs. We note that \cref{alg:bg} expects \cref{clock_addition_condition}, which will be defined later in \cref{sec:pipeline}, is satisfied for an $\ii$ implementation.

\subsubsection{Read-Modify-Write Removal on BRAM}
\begin{figure}[tbp]
    \centering
    \includegraphics[width=0.8\linewidth,pagebox=mediabox]{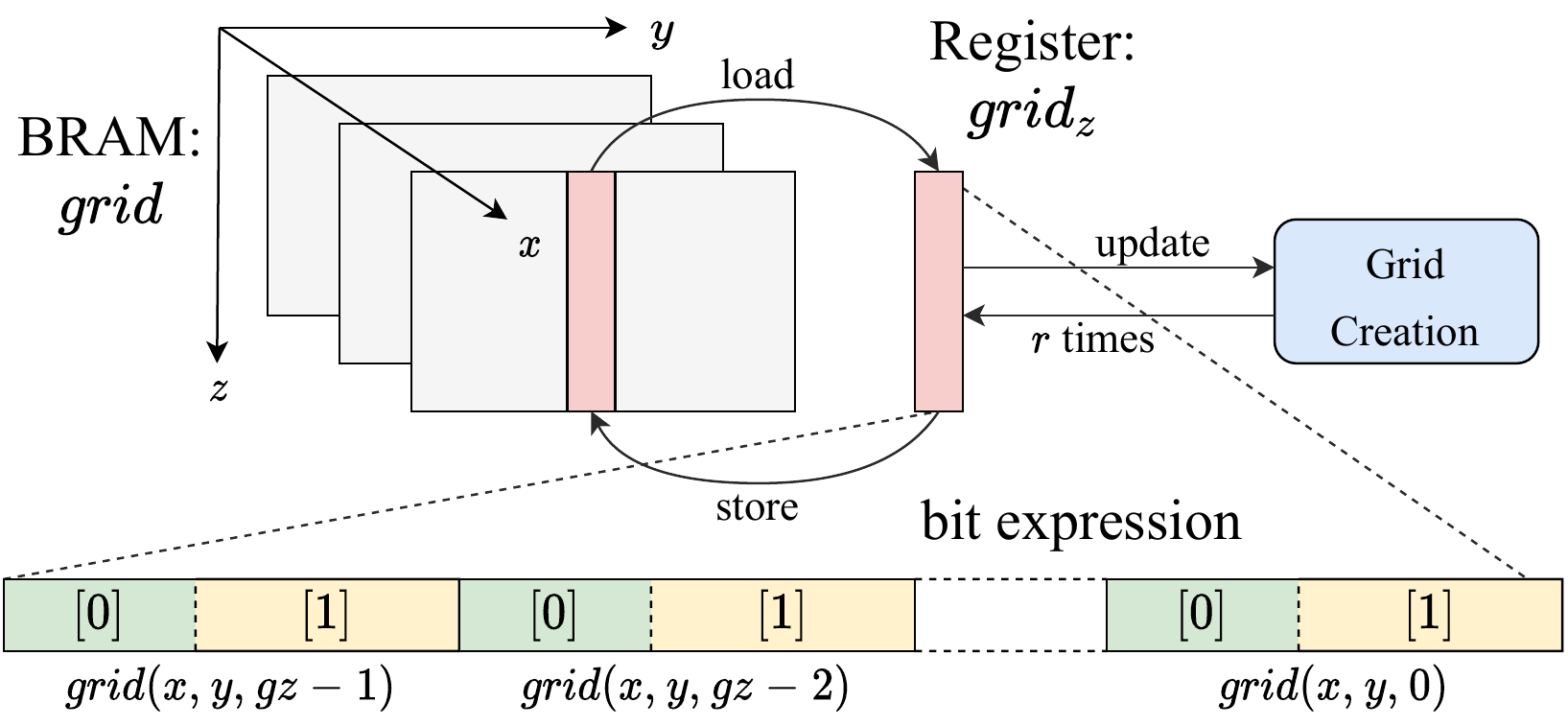}
    \caption{Illustration of read-modify-write removal and bit expression of $\gridds(x, y)$.}
    \label{fig:read_modify_write}
\end{figure}
For real-time processing, an FPGA design whose II (Initiation Interval) is 1 is desired. However, because $grid$ is stored in BRAMs, the read-modify-write operation $grid([\bfv(\bi)]) \ape (1, f(\bi))$ in the GC cannot be achieved in $\ii$. Here, by exploiting the characteristics that the accesses to $grid$ are not random but regular and local to some extent, this operation can be accelerated. The x- and y-components of $\bfv(\bi)$ change regularly as input values are read, and therefore, they can be expressed by counters ($\ell$. 36 to 45 in \cref{alg:bg}). However, the z-component of $\bfv(\bi)$ remains unknown until $f(\bi)$ is read. Therefore, as shown in \cref{fig:read_modify_write}, to update the elements $grid(x, y, \ast)$, the best solution would be to load $grid(x, y, \ast)$ onto registers $grid_z$ when processing the first pixel of each row in $gg(x, y)$ ($\ell$. 12 to 15 in \cref{alg:bg}), update them on registers $grid_z$ ($\ell$. 16 in \cref{alg:bg}), and store them back to BRAMs after processing the last pixel of the row in $gg(x, y)$ ($\ell$. 17 to 18 in \cref{alg:bg}). We note that the update operation on registers can be achieved in $\ii$. In this manner, the read-modify-write operation is removed.

\begin{figure}[h!]
\centering
\begin{algorithm}[H]
\begin{algorithmic}[1]
    \caption{Calculation of the BG on an FPGA.}
    \label{alg:bg}
    \Require{$\bfo$: input image}
    \Ensure{$\bfbg$: output image filtered by the BG}
    \State // Initialization
    \State $cx, py, cy \leftarrow r \am [r/2] \am 1, 0, r \am [r/2] \am 1$
    \State $\gridds[*][*] \leftarrow 0$
    \For{$x \leftarrow 0$ to $(h \am 1) \ap 2r \ap [r/2]$}
        \For{$y \leftarrow 0$ to $w \am 1$}
            \State // GC (while there are still input pixels)
            \If{$x < h$}
                \State $l \leftarrow \bfo(x, y)$
                \State $pz \leftarrow L_1[l]$
                \State $gl \leftarrow (1, l)$
                \State $lb$.enqueue($l$)
                \If{$(x,y)$ is the upper left corner of any $gg$}
                    \State $grid_z \leftarrow 0$
                \ElsIf{$(x,y)$ is the left end of any $gg$}
                    \State $grid_z \leftarrow \gridds[2][py]$
                \EndIf
                \State $grid_z[pz] \leftarrow grid_z[pz] \ap gl$
                \If{$(x,y)$ is the right end of any $gg$}
                    \State $\gridds[2][py] \leftarrow grid_z$
                \EndIf
            \EndIf
            \State // GF ($gy \at gz$ times for each plane
            \State // after necessary data is prepared)
            \If{$\gridds[2][1]$ completed}
                \State $cnt_y, cnt_z \leftarrow 0, 0$
            \EndIf
            \If{$\neg (cnt_y == gy \land cnt_z == 0)$}
                \State $gf[cnt_z] \leftarrow \fgf(2, cnt_y, cnt_z)$
                \If{$cnt_z == gz \am 1$}
                    \State $\gfds[1][cnt_y] \leftarrow gf$
                    \State Shift $\gridds[*][cnt_y+2]$ and $reg_{GF}$
                    \State $cnt_y, cnt_z \leftarrow cnt_y \ap 1, 0$
                \Else
                    \State $cnt_z \leftarrow cnt_z \ap 1$
                \EndIf
            \EndIf
            \State // TI (after two planes of $\gfds$ completed)
            \If{$x \geq 2r \ap [r/2]$}
                \State $\bfbg(x \am 2r \am [r/2], y)$
                \State \quad $\leftarrow \fti(L_2[cx], L_3[cy], lb$.dequeue())
                \State Load or Shift $\gfds$ and $reg_{TI}$ if necessary
            \EndIf
            \State // Update counters
            \If{$y == w \am 1$}
                \If{$cx == r \am 1$}
                    \State $cx, py, cy \leftarrow cx \ap 1, 0, r \am [r/2] \am 1$
                \Else
                    \State $cx, py, cy \leftarrow 0, 0, r \am [r/2] \am 1$
                \EndIf
            \ElsIf{$cy == r \am 1$}
                \State $py, cy \leftarrow py \ap 1, 0$
            \Else
                \State $cy \leftarrow cy \ap 1$
            \EndIf
        \EndFor
    \EndFor
\end{algorithmic}
\end{algorithm}
\end{figure}

The suitable data structure for the grid is a 2D space $\gridds$ with x- and y-axes, because all elements in the z-axis direction with a certain x- and y-components must be loaded and stored together. We note that each $(x, y)$ element in $\gridds$ expresses $grid(x, y, \ast)$, which means that the value in $\gridds(x, y)$ is expressed as a bit combination of
\begin{align}
    & grid(x, y, gz-1)[0], grid(x, y, gz-1)[1], \\
    & \quad grid(x, y, gz-2)[0], grid(x, y, gz-2)[1], \\
    & \qquad \ldots, grid(x, y, 0)[0], \textrm{ and } grid(x, y, 0)[1]
\end{align}
in this order (see \cref{fig:read_modify_write}). Hereafter, the notation $\gridds(x, y)$ is used instead of $grid(x, y, \ast)$ to simplify the explanations and improve ease of understanding.

\subsubsection{Pipeline} \label{sec:pipeline}
\begin{figure}[tbp]
    \centering
    \includegraphics[width=0.9\linewidth,pagebox=mediabox]{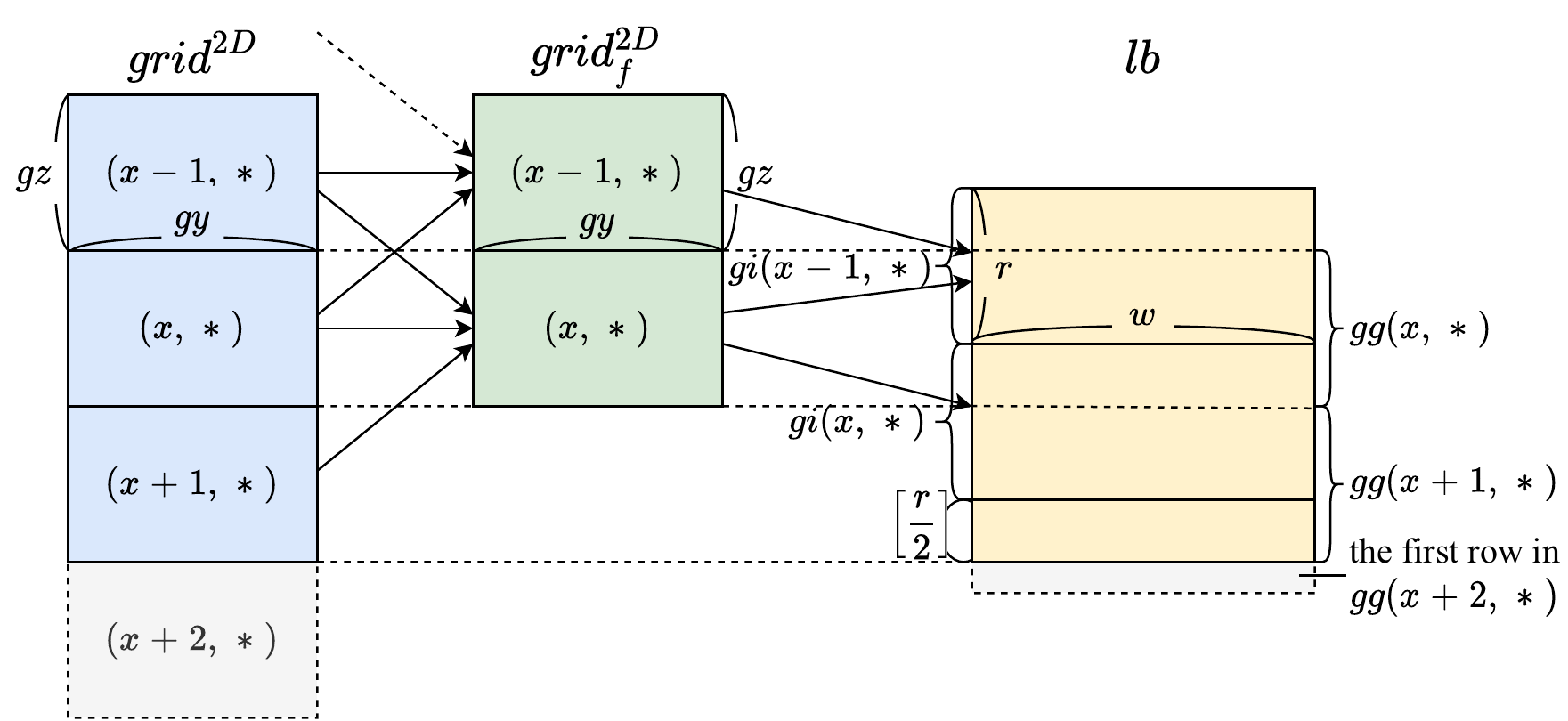}
    \caption{Data layout of $\gridds$, $\gfds$, and $lb$. The arrows represent data dependencies. The boxes surrounded by solid lines indicate the memory usage required for the proposed design.}
    \label{fig:additional_memory}
\end{figure}
\begin{figure}[tbp]
    \centering
    \begin{minipage}[b]{\linewidth}
        \centering
        \includegraphics[keepaspectratio, width=\linewidth,pagebox=mediabox, trim=0 0 60 0, clip]{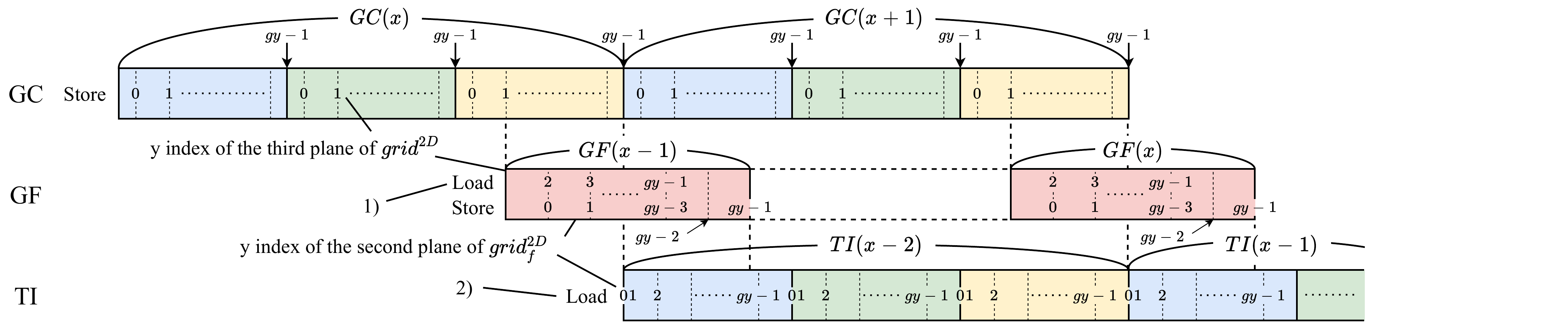}
        \subcaption{GF is sufficiently short.}
        \label{fig:pipeline-b}
    \end{minipage}\\
    \begin{minipage}[b]{\linewidth}
        \centering
        \includegraphics[keepaspectratio, width=\linewidth,pagebox=mediabox, trim=0 0 60 0, clip]{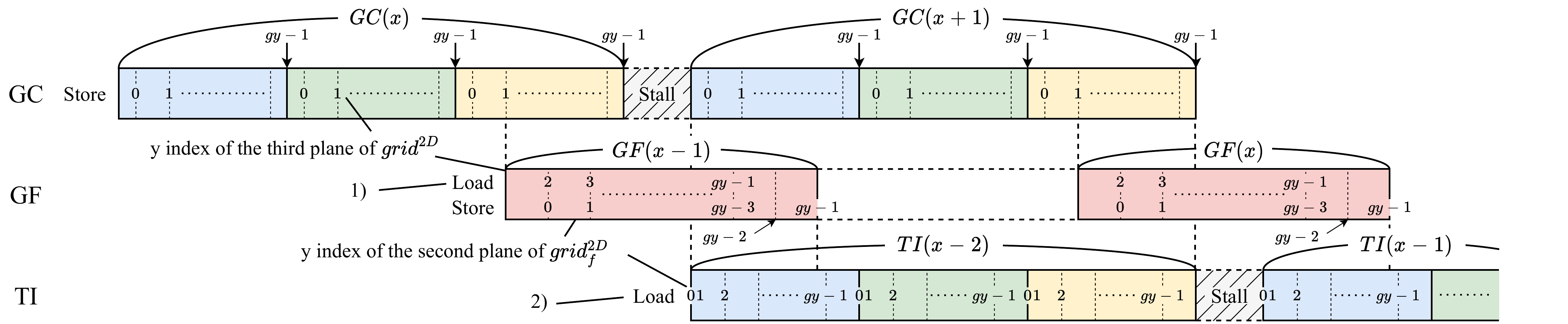}
        \subcaption{GF is sufficiently long.}
        \label{fig:pipeline-d}
    \end{minipage}
    \caption{Possible pipeline cases when $r \aeq 3$. $GC(x)$ expresses the GC to generate $\gridds(x, \ast)$; $GF(x)$ expresses the GF to generate $\gfds(x, \ast)$; and $TI(x)$ expresses the TI to generate $gi(x, \ast)$ in the output. 1) indicates that 0 and 1 are loaded onto $reg_{GF}$ directly from $\gridds$. 2) indicates that 1 can be loaded one clock after 0 because the first pixel can be interpolated by only 0.}
    \label{fig:pipeline}
\end{figure}

For further performance improvement, three for loops of the GC, GF, and TI are pipelined together, which means they are unified into one for loop. Because the GF is executed for each grid element and the GC and TI are executed for each input image pixel, the number of executions of the GF, $gx \at gy \at gz$, is different from that of the GC and TI, $w \at h$. Moreover, these processes are dependent on each other. Therefore, hardware-level pipeline of such heterogeneous processes is challenging. \cref{fig:additional_memory} shows the data dependencies and memory usage, which is much smaller than one whole image. $\gfds$ is defined in the same way as $\gridds$, and hereafter, $\gfds(x, y)$ is used instead of $\gf(x, y, \ast)$. Here, we focus on the GF of the plane $\gridds(x, \ast)$ in the grid. The nine lines around the line of interest should be loaded onto registers from BRAMs before the line is processed. Thus, this operation can start after the generation of the second line $\gridds(x \ap 1, 1)$. Furthermore, this operation should be completed before the last line $\gfds(x, gy \am 1)$ is loaded, which is required for the TI. Therefore, if
\begin{align}
    gy \times gz < 2 w - \round{\dfrac{r}{2}} - r - (w \bmod r),
    \label{clock_addition_condition}
\end{align}
holds, the GF can finish in time, as shown in \cref{fig:pipeline-b}. Otherwise, as shown in \cref{fig:pipeline-d}, the GC is delayed by suspending the input until it can restart. The TI is then processed $(r \ap \left[ \frac{r}{2} \right])$ lines behind the GC. In this manner, the pipeline between the set of processes ($GC(x)$, $GF(x-1)$, and $TI(x-2)$) are designed, which is called macro pipeline, and at the same time, the pipeline within the set of processes are also performed, which is called micro pipeline. This nested pipeline structure greatly accelerates our design.

\subsubsection{Other Optimizations}
\begin{figure}[tbp]
    \centering
    \includegraphics[width=0.6\linewidth,pagebox=mediabox]{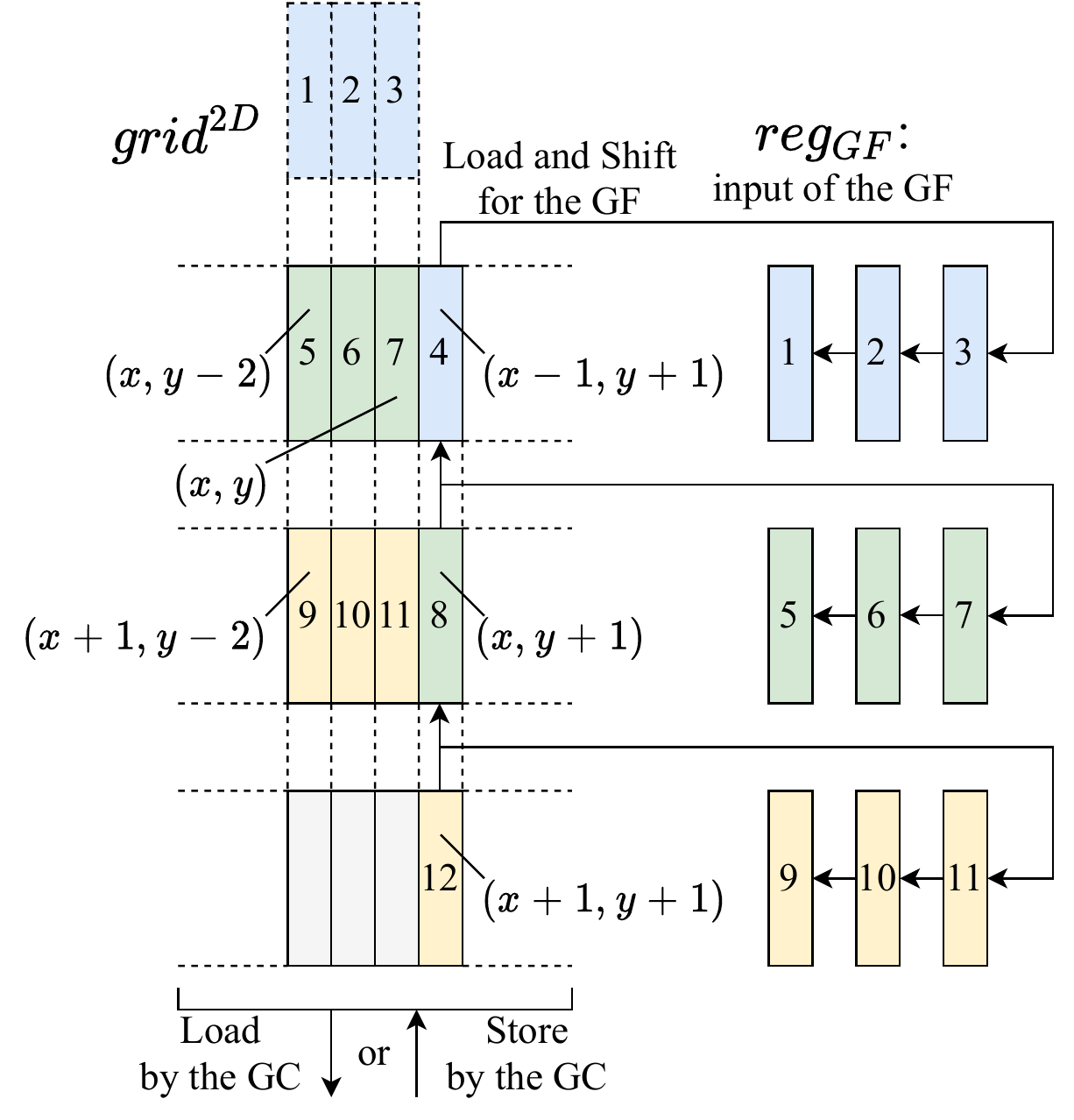}
    \caption{Accesses to $grid$ by the GC and GF. The direction of the arrows indicates where the data are loaded and stored. The numbers in $\gridds$ correspond to those in registers $reg_{GF}$. The dashed boxes (1, 2, and 3) no longer exist in $\gridds$. The GC does not access 12 in this situation.}
    \label{fig:grid_hazards}
\end{figure}
\begin{figure}[tbp]
    \centering
    \begin{minipage}[b]{0.49\linewidth}
        \centering
        \includegraphics[keepaspectratio, width=\linewidth,pagebox=mediabox]{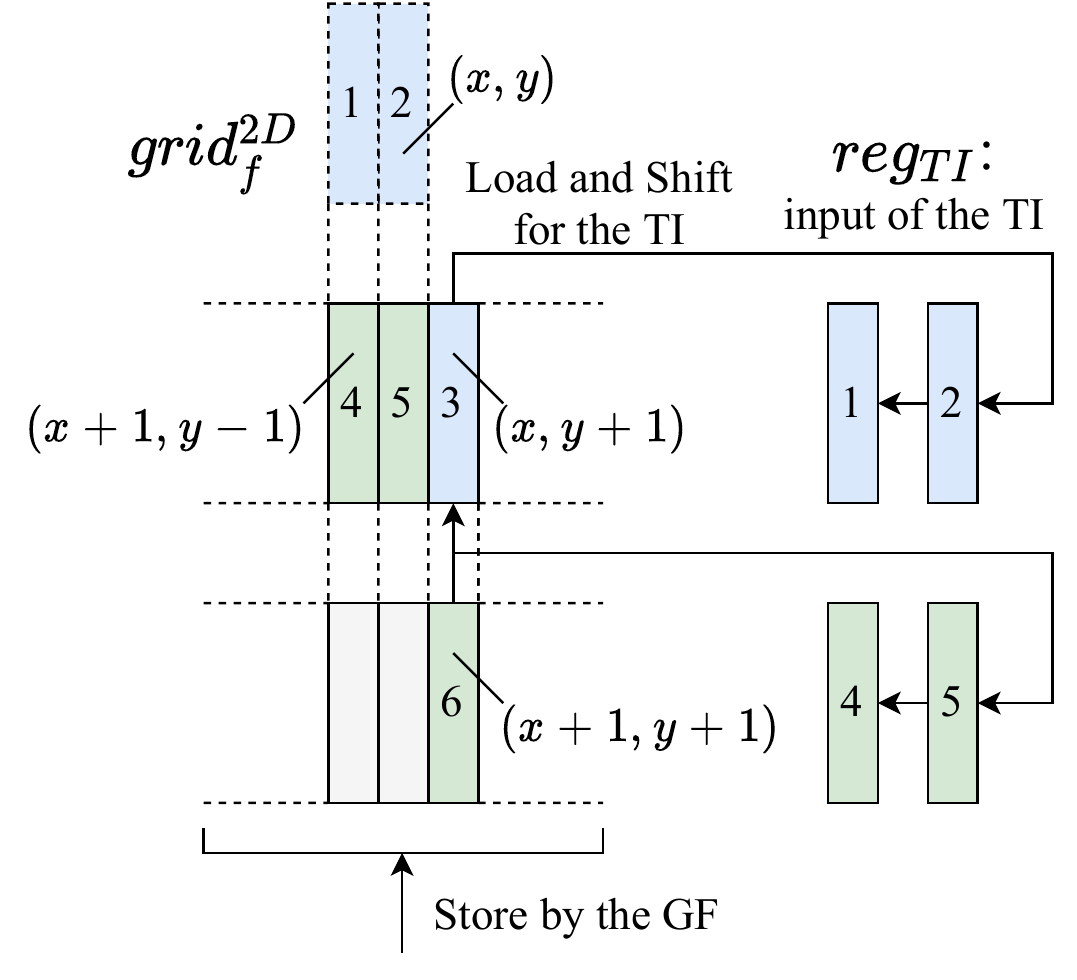}
        \subcaption{Scenario where the last row in $gi(x, y)$ is processed.}
        \label{fig:gf_hazards-b}
    \end{minipage}
    \begin{minipage}[b]{0.49\linewidth}
        \centering
        \includegraphics[keepaspectratio, width=\linewidth,pagebox=mediabox]{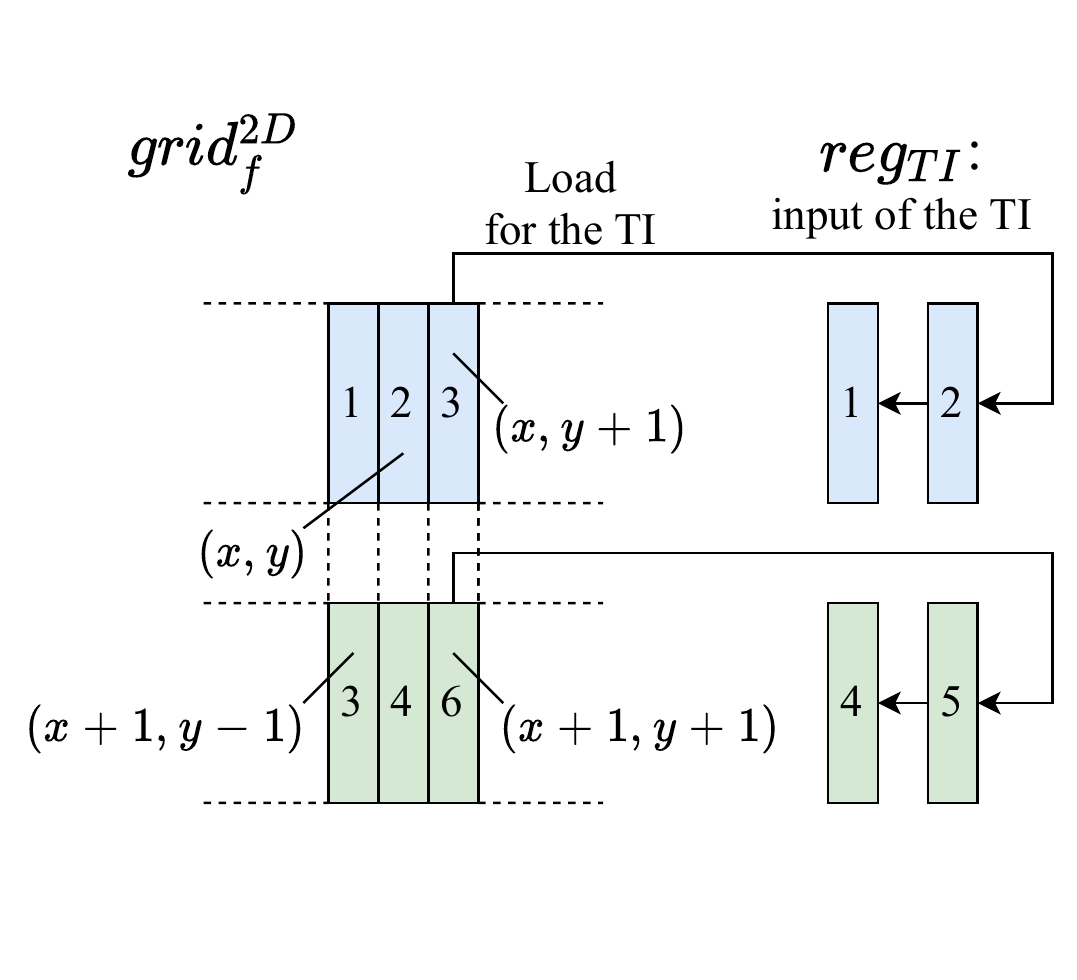}
        \subcaption{Scenario where the rest of the rows in $gi(x, y)$ are processed.}
        \label{fig:gf_hazards-a}
    \end{minipage}
    \caption{Accesses to $\gf$ by the GF and TI. The direction of the arrows indicates where the data are loaded and stored. The numbers in $\gfds$ correspond to those in registers $reg_{TI}$. The dashed boxes (1 and 2 in (a)) no longer exist in $\gfds$. The GF does not access 6 in this situation.}
    \label{fig:gf_hazards}
\end{figure}
We also partition the BRAMs to remove structural hazards. At most two load and $\ad$ or store operations are allowed in one BRAM in one clock. The accesses to $grid$ and $\gf$ are shown in \cref{fig:grid_hazards,fig:gf_hazards}, respectively. Therefore, $grid$ and $\gf$ should be partitioned into three and two by plane such that each partition is not accessed more than twice. In terms of $lb$, because the input pixels are directly stored and the stored data are loaded for the TI in the same order, $lb$ is implemented as FIFO without causing structural hazards. Therefore, all structural hazards are removed, and $\ii$ implementation is achieved in the proposed design.

\begin{figure}[tbp]
    \centering
    \includegraphics[width=\linewidth,pagebox=mediabox]{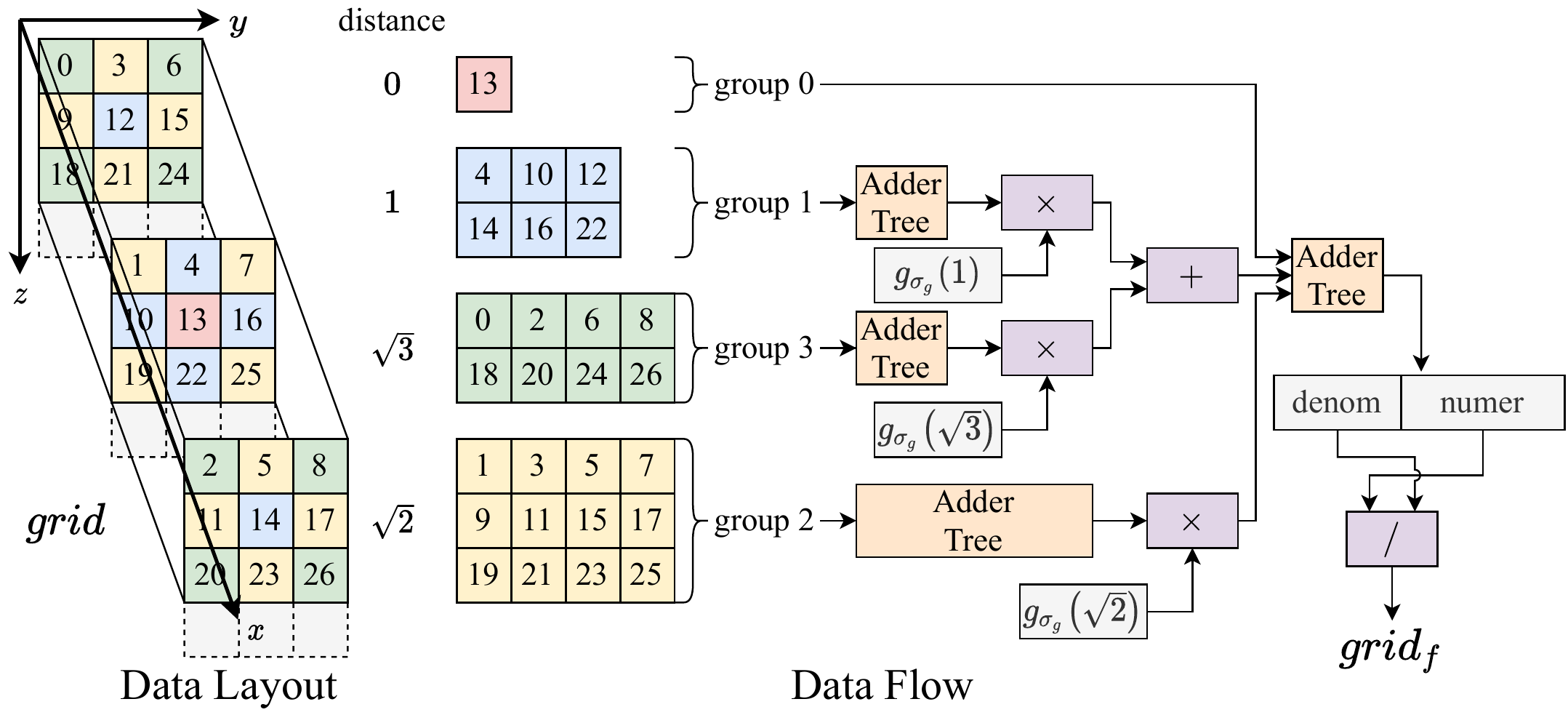}
    \caption{Illustration of the GF calculations.}
    \label{fig:gaussian}
\end{figure}
\begin{figure}[t!]
    \centering
    \includegraphics[width=0.7\linewidth,pagebox=mediabox]{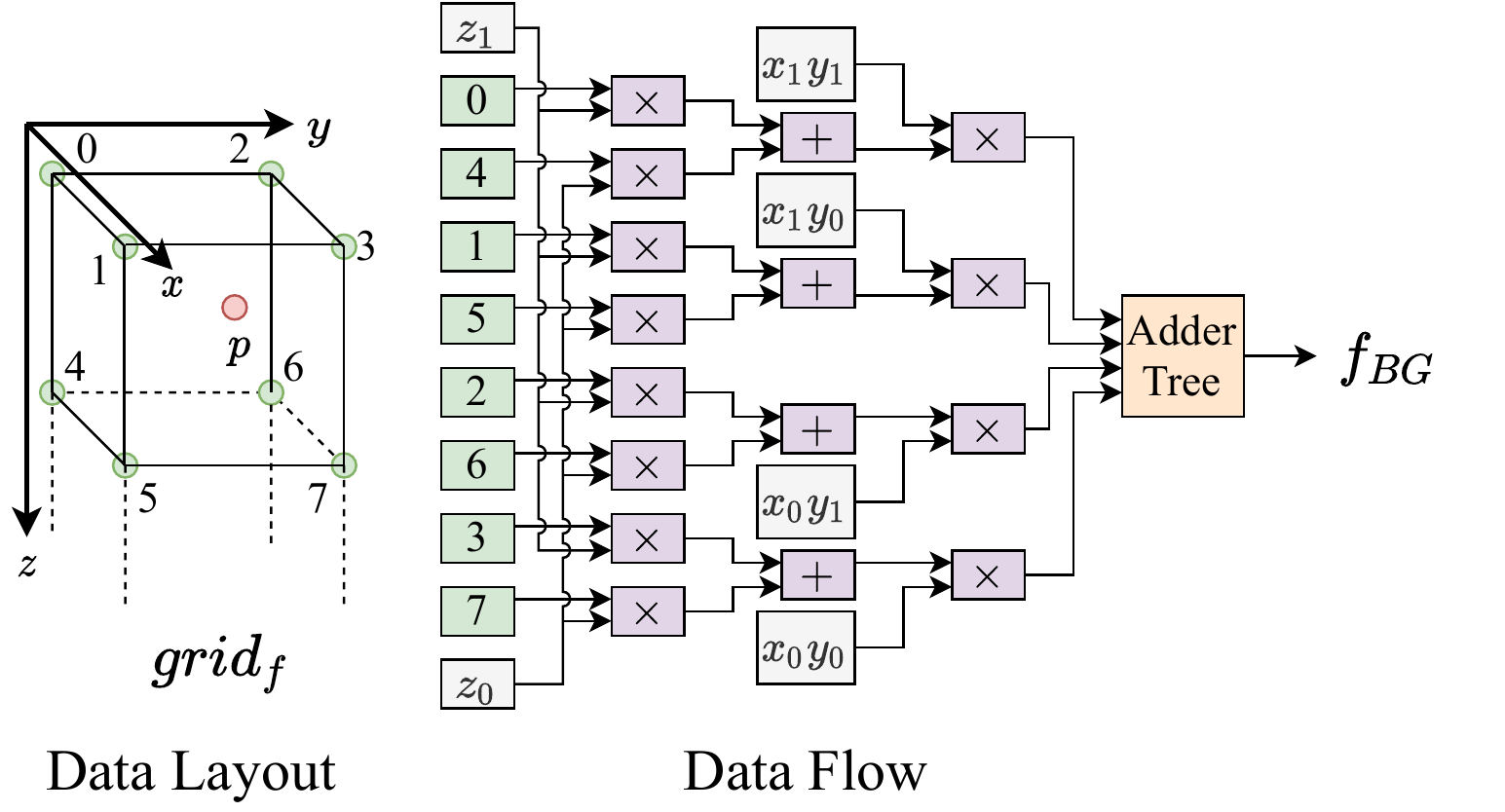}
    \caption{Illustration of the TI calculations.}
    \label{fig:interpolation}
\end{figure}
Finally, the arithmetic units are minimized. To remove floating-point arithmetic, we utilize a simple approach to multiply each value by a power of two, which can be implemented as shift operations. Then, the GF \cref{gf3D} and TI \cref{interpolate} can be calculated as shown in \cref{fig:gaussian,fig:interpolation}, respectively. We note that the numerator and denominator in the GF can be calculated together owing to the data structure of the $grid$.

\section{Implementation and Evaluation} \label{sec:implementation_and_evaluation}
For comparison, the proposed design is implemented on the ZCU 104 board with Zynq UltraScale+ MPSoC XCZU7EV-2FFVC1156 from Xilinx using Vivado HLS 2019.2 and Vivado 2019.2. Then, we run the implemented design using PYNQ v2.6 for the ZCU 104 board.

\subsection{Denoising Quality}
\newcommand{\signval}{30}
\newcommand{\sigsval}{4}
\newcommand{\sigrval}{50}
\newcommand{\rval}{7}
\begin{figure}[tbp]
    \centering
    \begin{minipage}[b]{0.49\linewidth}
        \centering
        \includegraphics[keepaspectratio, width=\linewidth]{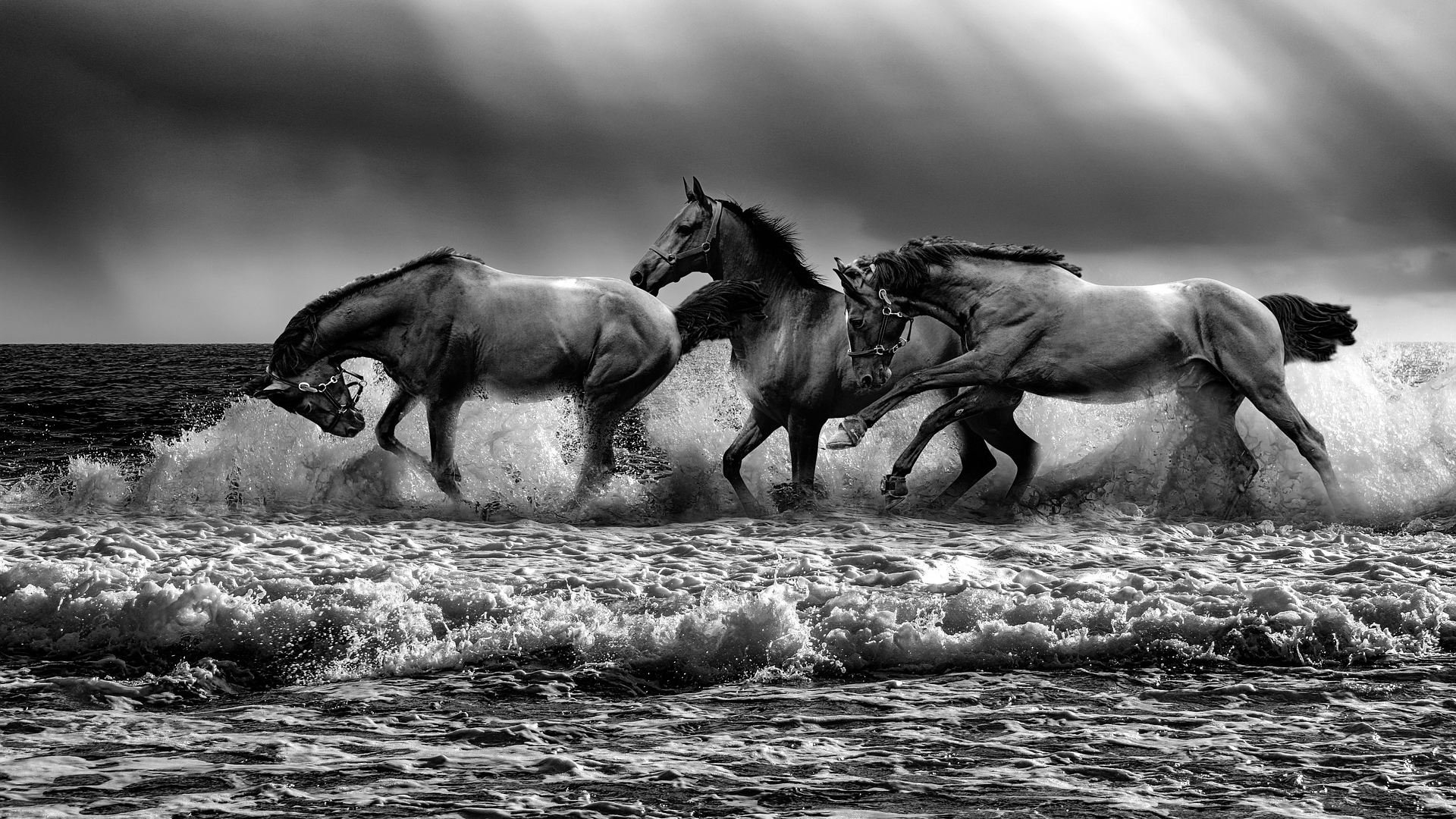}
        \subcaption{Original}
        \label{fig:horse}
    \end{minipage}
    \begin{minipage}[b]{0.49\linewidth}
        \centering
        \includegraphics[keepaspectratio, width=\linewidth]{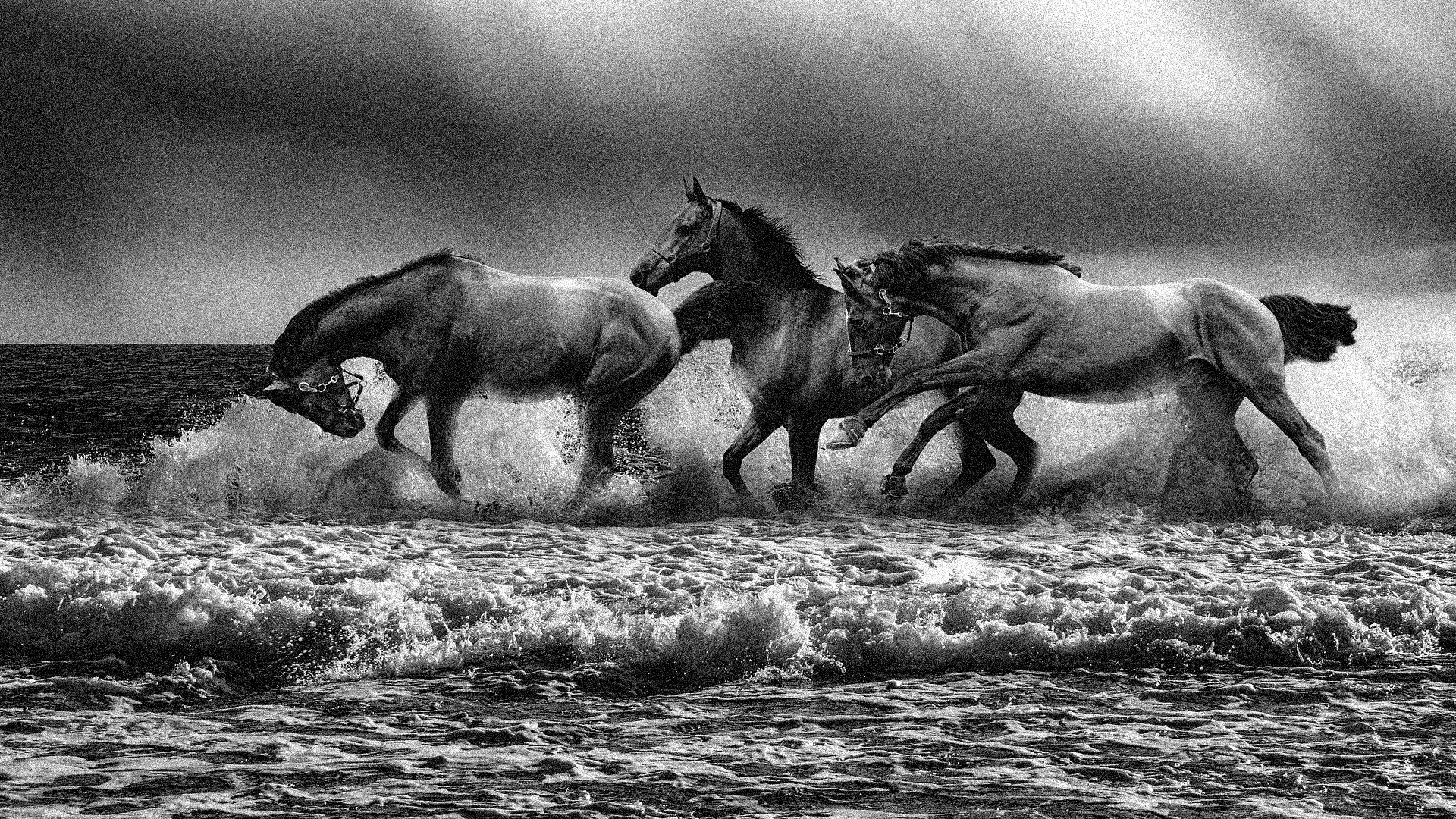}
        \subcaption{Noised}
        \label{fig:noised_horse}
    \end{minipage}
    \caption{Horse pictures used for evaluation.}
    \label{fig:pictures}
\end{figure}
\begin{figure}[tbp]
    \centering
    \begin{minipage}[b]{0.49\linewidth}
        \centering
        \includegraphics[keepaspectratio, width=\linewidth]{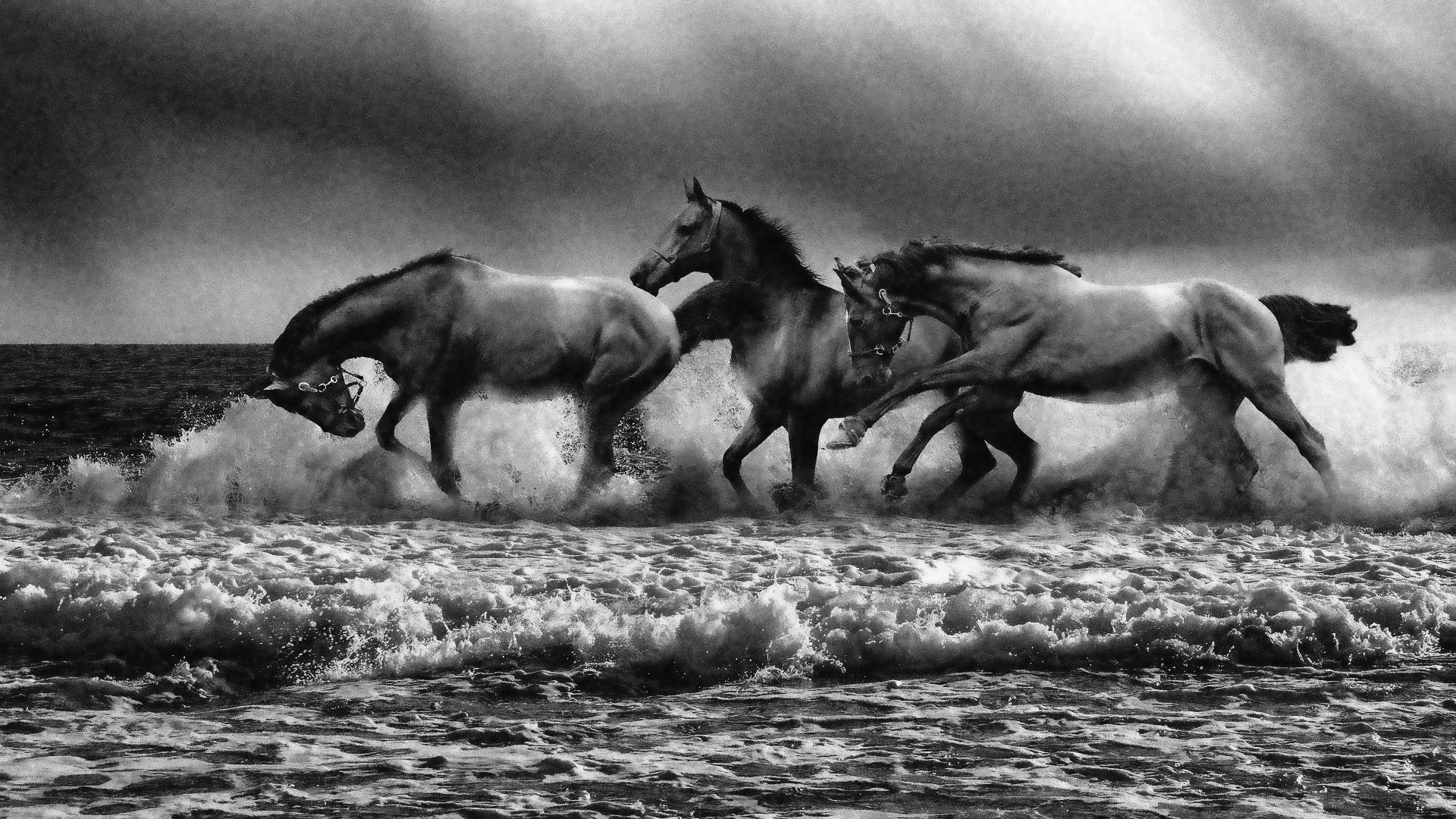}
        \subcaption{Denoised by the BF}
        \label{fig:denoised_horse_bf}
    \end{minipage}
    \begin{minipage}[b]{0.49\linewidth}
        \centering
        \includegraphics[keepaspectratio, width=\linewidth]{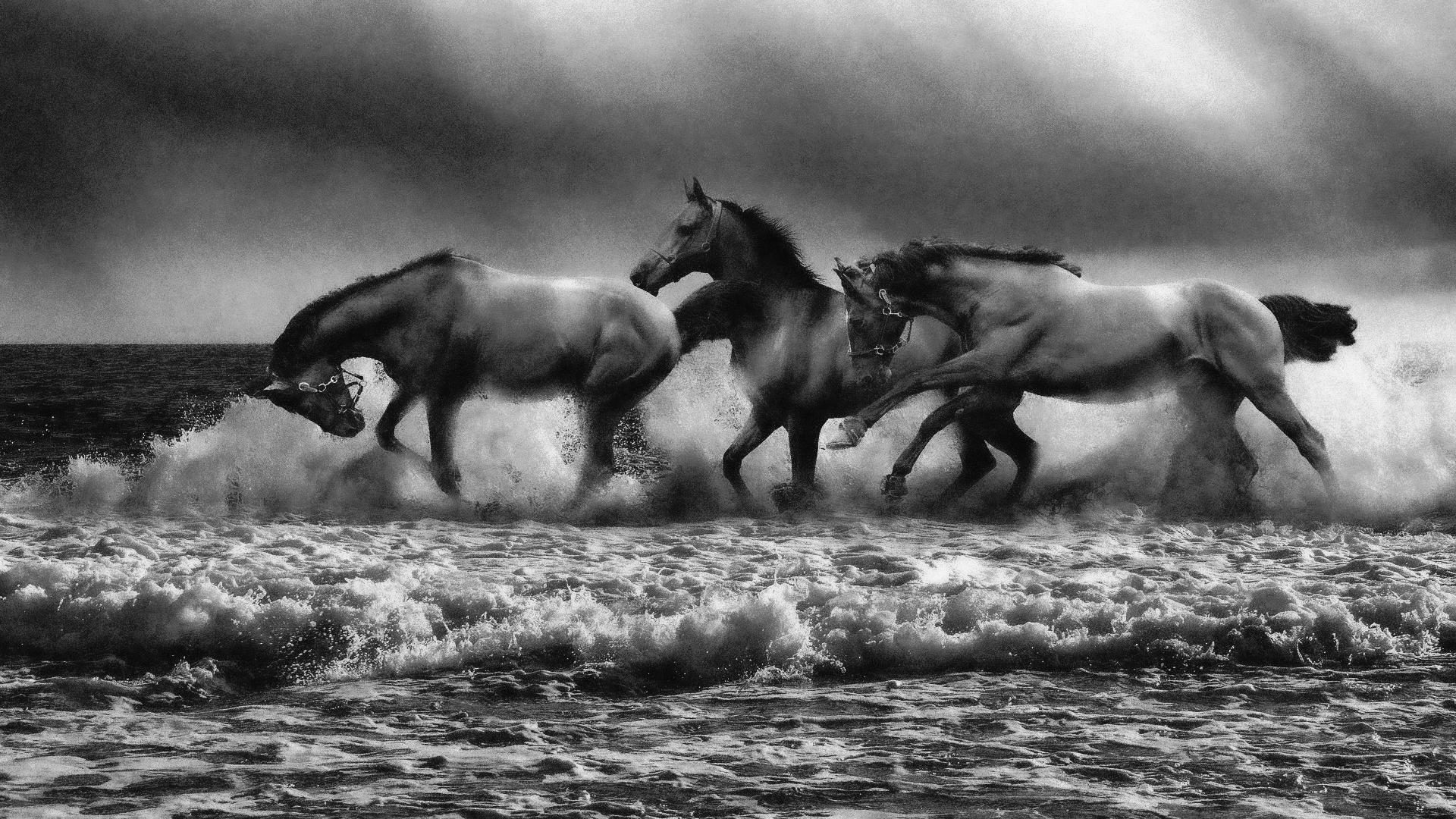}
        \subcaption{Denoised by the BG}
        \label{fig:denoised_horse_bg_fpga}
    \end{minipage}
    \caption{Comparison of horse pictures denoised using the two different filters.}
    \label{fig:denoised_pictures}
\end{figure}
\begin{figure}[t!]
    \centering
    \begin{minipage}[b]{0.24\linewidth}
        \centering
        \includegraphics[keepaspectratio, width=\linewidth]{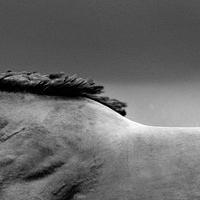}
        \subcaption{Original}
        \label{fig:horse_clip}
    \end{minipage}
    \begin{minipage}[b]{0.24\linewidth}
        \centering
        \includegraphics[keepaspectratio, width=\linewidth]{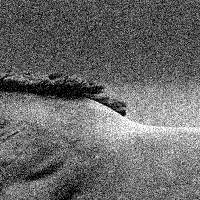}
        \subcaption{Noised}
        \label{fig:noised_horse_clip}
    \end{minipage}
    \begin{minipage}[b]{0.24\linewidth}
        \centering
        \includegraphics[keepaspectratio, width=\linewidth]{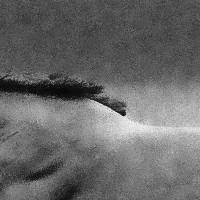}
        \subcaption{BF}
        \label{fig:denoised_horse_bf_clip}
    \end{minipage}
    \begin{minipage}[b]{0.24\linewidth}
        \centering
        \includegraphics[keepaspectratio, width=\linewidth]{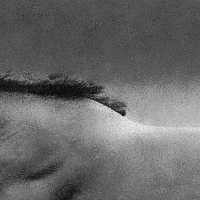}
        \subcaption{BG}
        \label{fig:denoised_horse_bg_fpga_clip}
    \end{minipage}
    \caption{Trimmed pictures of \cref{fig:pictures,fig:denoised_pictures} for comparison.}
    \label{fig:pictures_comparison}
\end{figure}
Here, we evaluate the denoising quality by using two filters: the BF and the proposed BG on an FPGA. We use a full HD ($(w, h) \aeq (1920, 1080)$) grayscale picture (see \cref{fig:pictures}) to clarify that our design can process large-scale and high-resolution images. To measure the denoising quality, which is defined by the extent to which denoised pictures using the filters (\cref{fig:denoised_pictures}) differ from the original picture (\cref{fig:horse}), we use the criterion MSSIM (Mean Structural SIMilarity index) proposed by \cite{ssim}. The MSSIM corresponds to human visual perception to a higher degree than the other criteria, such as the PSNR (Peak Signal-to-Noise Ratio). The hyperparameters $C_1$ and $C_2$ are fixed as $(0.01 \at 255)^2$ and $(0.03 \at 255)^2$, respectively, and $7 \at 7$ square window is used (see \cite{ssim} for more details). The denoising quality is better if the MSSIM value is larger, and the maximum value is 1.0. If the value is 1.0, the two pictures are identical.

First, from the original picture (\cref{fig:horse}), the noised picture (\cref{fig:noised_horse}) is created by adding Gaussian noise with a standard deviation of \signval. Then, the noised picture is processed using the two filters to obtain denoised pictures (\cref{fig:denoised_pictures}). Finally, we calculate the MSSIM values between the denoised pictures and the original picture. The results are obtained by changing $r$, $\sigr$, and $\sigs$.

\begin{figure}[tbp]
    \centering
    \begin{minipage}[b]{0.32\linewidth}
        \centering
        \includegraphics[keepaspectratio, width=\linewidth]{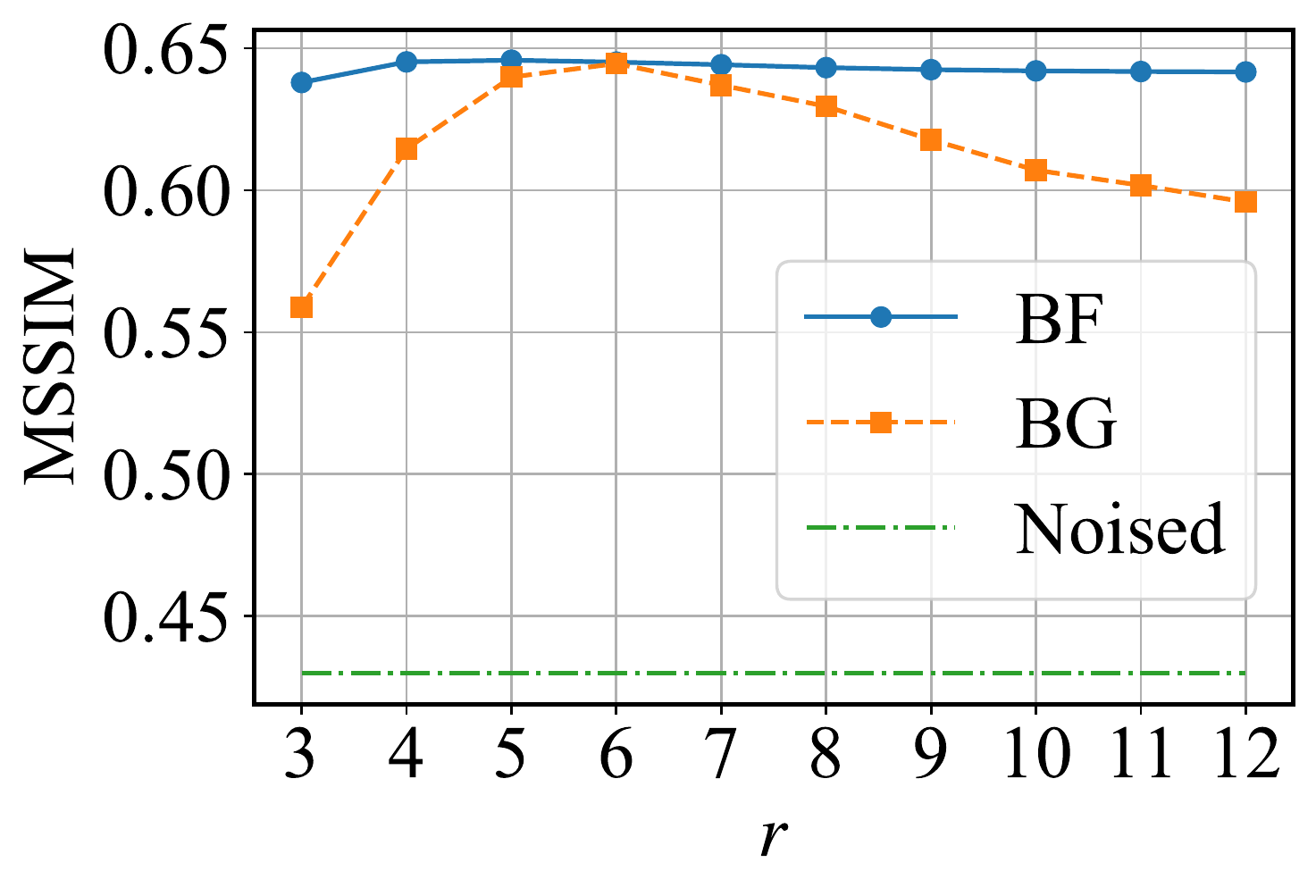}
        \subcaption{$(\sigs, \sigr) \aeq (\sigsval, \sigrval)$}
        \label{fig:quality_horse_r}
    \end{minipage}
    \begin{minipage}[b]{0.32\linewidth}
        \centering
        \includegraphics[keepaspectratio, width=\linewidth]{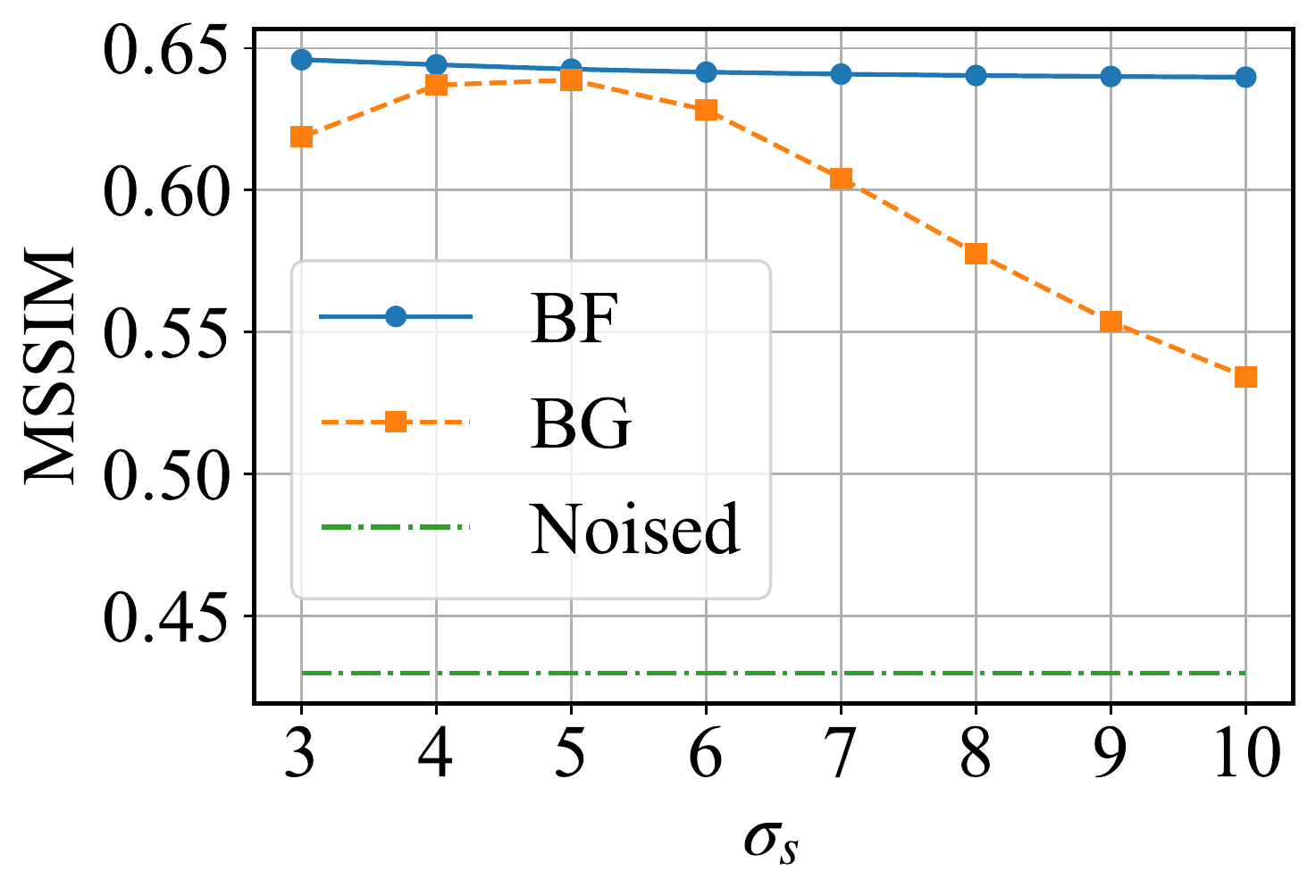}
        \subcaption{$(r, \sigr) \aeq (\rval, \sigrval)$}
        \label{fig:quality_horse_ss}
    \end{minipage}
    \begin{minipage}[b]{0.32\linewidth}
        \centering
        \includegraphics[keepaspectratio, width=\linewidth]{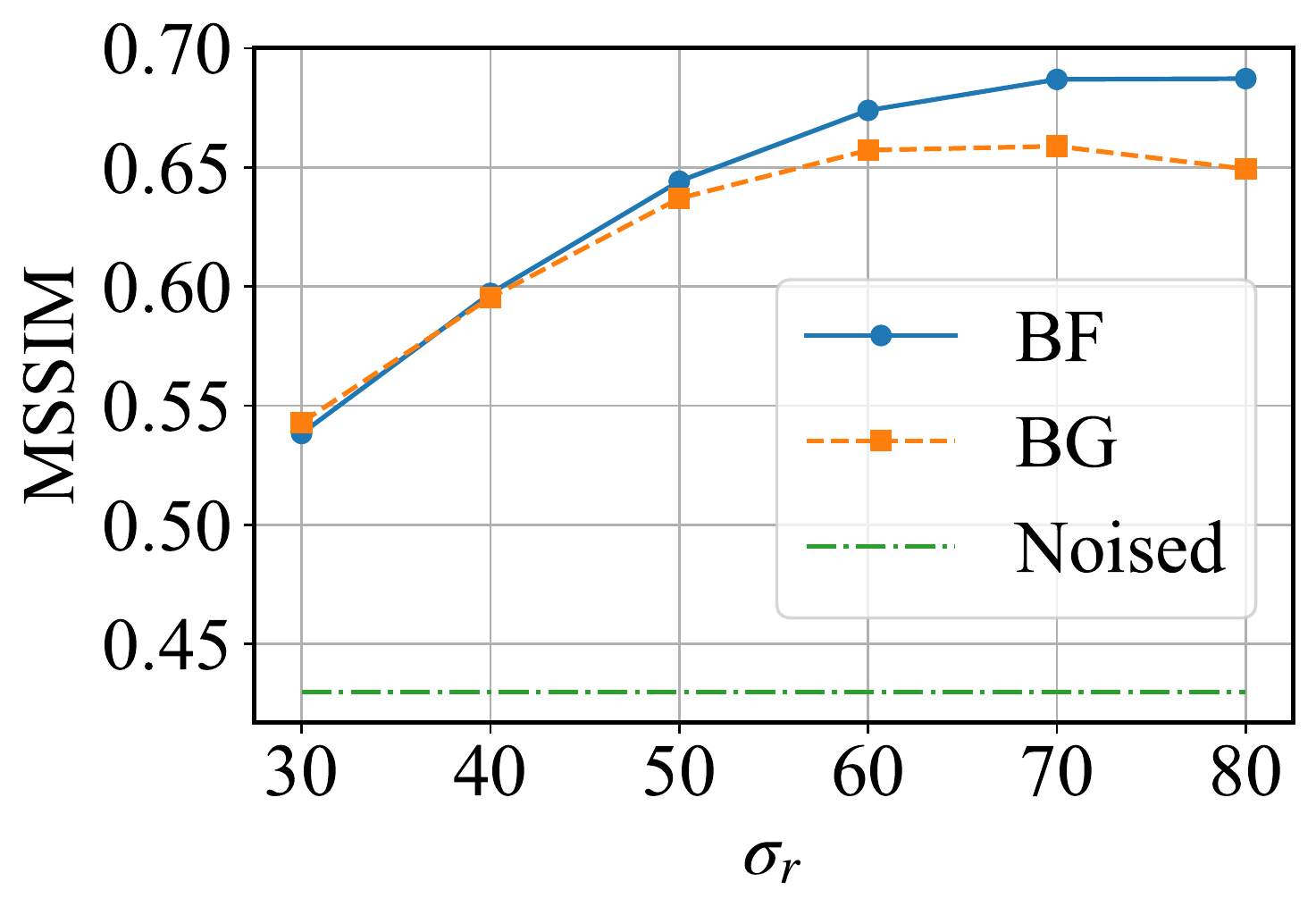}
        \subcaption{$(r, \sigs) \aeq (\rval, \sigsval)$}
        \label{fig:quality_horse_sr}
    \end{minipage}
    \caption{Relationship between several parameters and the MSSIM values of the pictures before and after denoising.}
    \label{fig:quality_horse}
\end{figure}
As shown in \cref{fig:quality_horse}, the MSSIM values obtained from the BF are larger than those obtained from the BG. However, by selecting proper parameters, the BG shows equivalent denoising quality in terms of the MSSIM. Moreover, \cref{fig:pictures_comparison} indicates that the BG shows equal or better denoising quality compared to the BF.

\subsection{Computation Speed and Hardware Resources}
\begin{table}[tbp]
    \centering
    \caption{Comparison of the speed and resources of the proposed design by changing $r$ when $\sigr \aeq 70$ and $\sigs \aeq 8$.}
    \includegraphics[width=0.9\linewidth,pagebox=mediabox]{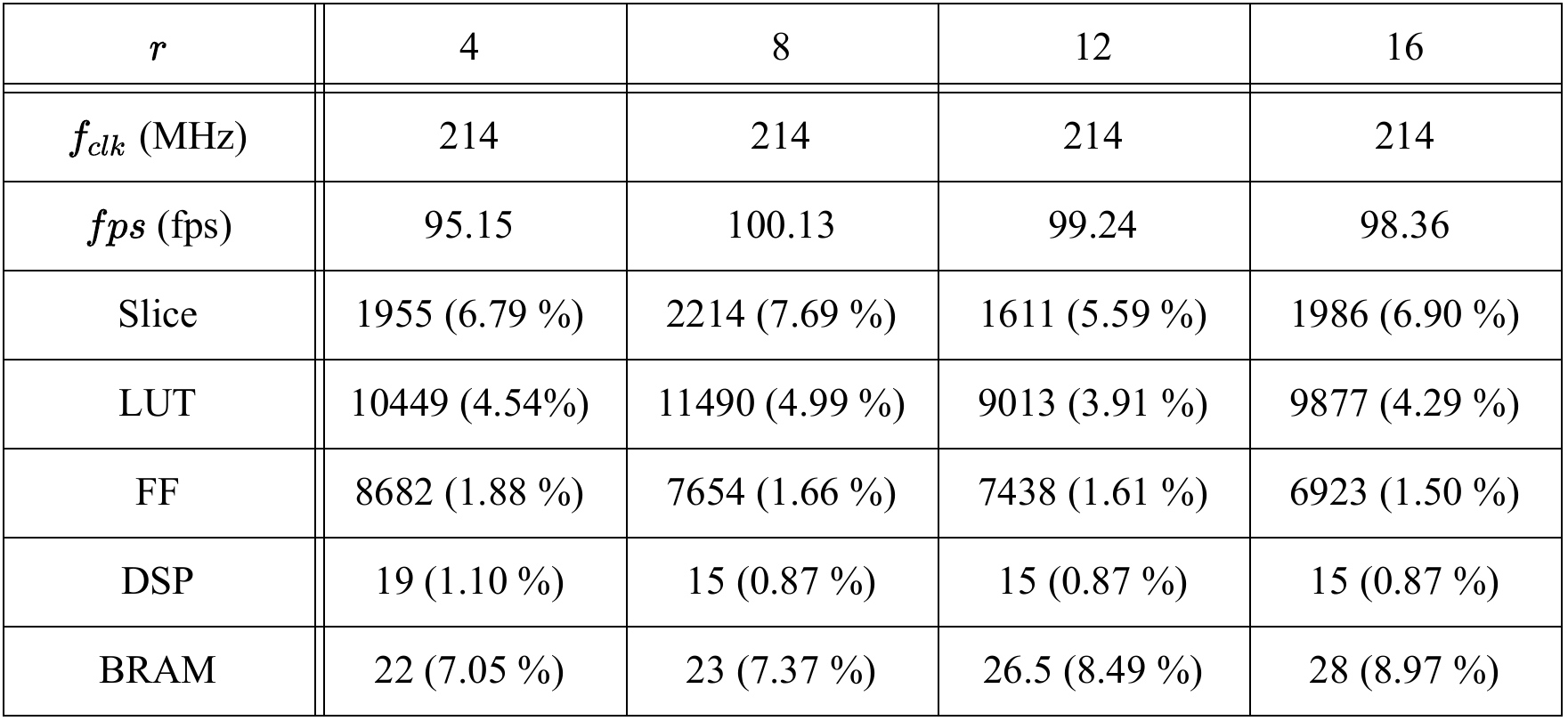}
    \label{tb:proposed_evaluation}
\end{table}
First, the computation speed and hardware resources of the proposed design are evaluated as shown in \cref{tb:proposed_evaluation}, where $f_{clk}$ and $fps$ denote the maximum clock frequency and actual maximum frame rate, respectively. The maximum clock frequency $f_{clk}$ is obtained by finding the frequency at which all timing constraints are met in the Vivado. The actual frame rate $fps$ is obtained by measuring the execution time per frame $T_f \aeq 1 / fps$ on the ZCU 104 board. The $fps$ values are almost the same as the theoretical values. The rest of the items are obtained from the actual implementation results in the Vivado.

In the proposed design, as shown in \cref{tb:proposed_evaluation}, the computation speed is sufficiently high in full HD images, and the speed is almost the same and independent of $r$. However, when $r$ is 4, the design runs slightly slowly because \cref{clock_addition_condition} does not hold, and extra clocks are required to finish the GF. Further, it is inferred that the consumption of hardware resources remains almost the same when $r$ increases.

Next, we compare the speed and resources of our design, a GPU implementation of the BF, and other existing implementations: (1) ICCEE 2008 \cite{brute_force_polynomial}, (2) TIE 2014 \cite{fully_synchronized}, and (3) TIE 2018 \cite{reconfigurable}. Our design and (3) have the characteristics wherein the consumption of hardware resources does not increase when $r$ is enlarged; however, (1) and (2) do not have the characteristics.

\begin{table}[tbp]
    \centering
    \caption{Comparison of speed and resources between our design, GPU implementation of the BF, and other existing implementations. 1) indicates that the value is the number of logic elements in Altera, which do not exactly correspond to the slices in Xilinx.}
    \includegraphics[width=\linewidth,pagebox=mediabox]{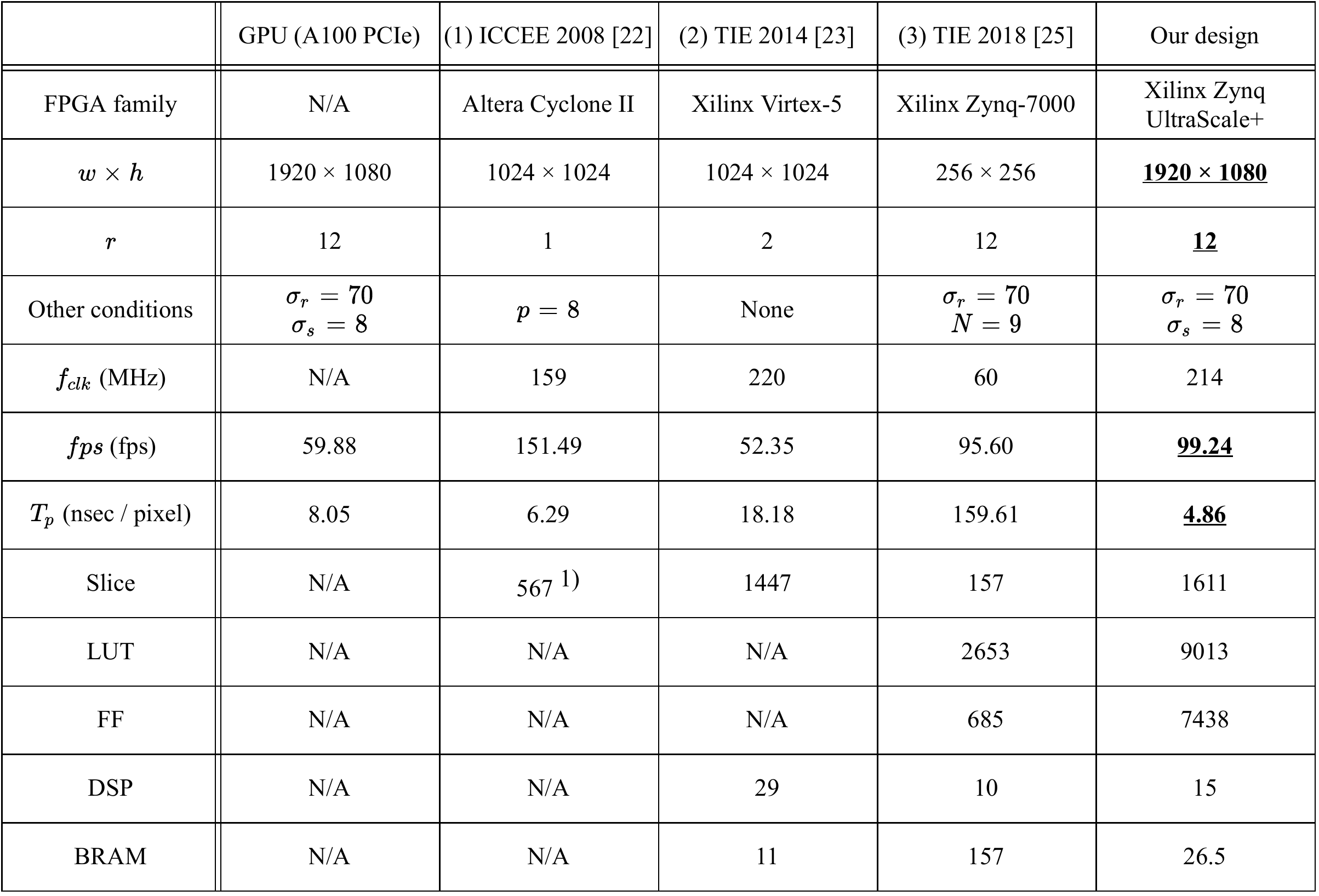}
    \label{tb:resource_speed}
\end{table}

The results are shown in \cref{tb:resource_speed}, where $f_{clk}$ and $T_p$ denote the maximum clock frequency and elapsed time per pixel, respectively. Here, we use estimated values for the frame rates of (1) to (3) (refer to the original papers for more details), because the actual values are not shown. As for the GPU implementation, we use one of the highest performance GPUs: A100 PCIe from NVIDIA. We also use the cv::cuda::bilateralFilter function in OpenCV 4.5.1 for C++ implementation and g++ 9.3.0 as a compiler.

As most of the filters are implemented on different FPGA boards, parameters, and sizes of images, the comparison of the speed of these implementations may be less significant. However, several insights can be obtained from these results. The elapsed time per pixel suggests that our design is the fastest of the five implementations, at least in this scenario, and ours is reasonably fast for real-time processing of large-scale and high-resolution images. In contrast, (1) can also process images relatively fast, but the resources increase in proportion to the square of $r$; (2) and (3) are much slower than our design.

As there are N/A cells in the table, the comparison of hardware resources may be incomplete. However, the results suggest that our design consumes a small number of BRAMs and DSPs, even though $r$ is large. In contrast, slice, LUT, and FF usage are not small compared with (3), but considering that our design runs sufficiently fast for real-time processing, it is acceptable because our design consumes a small percentage of resources on the ZCU 104 board, which is a relatively small-scale board. These outstanding characteristics are the result of highly parallelized and deeply pipelined implementation.

\section{Conclusion} \label{sec:conclusion}
In this paper, we provide a detailed explanation of the BG with a variable-sized window and its fully pipelined FPGA implementation. The advantages of the proposed design are summarized as follows.
\begin{enumerate}
    \item The BG is enhanced so that the window size of input images can be varied.
    \item The fully pipelined FPGA implementation is proposed for the proposed BG so that it can suppress the increase in the hardware resources.
    \item The proposed design is implemented on an actual FPGA board, and it outperforms the other existing designs in terms of computation speed and hardware resources.
\end{enumerate}

Moreover, there is some room for improvement in the proposed design, especially in terms of the sensitivity of its output to the variations in the parameters used and its application to higher memory bandwidth. This sensitivity can be reduced by further enhancing the BG algorithm with a variable-sized window. Furthermore, the adverse effects caused by sensitivity can be alleviated by selecting the best parameters in terms of their MSSIM values. Here, a higher memory bandwidth implies that more than one pixel is read and processed together. Therefore, the implementation requires some changes, although the basic theory remains the same.

\section*{Acknowledgment}
\addcontentsline{toc}{section}{Acknowledgment}
This work is supported in part by JSPS KAKENHI 19H04075 and 18H05288, and JST PRESTO JPMJPR18M9.



\end{document}